\documentclass[10pt, oneside, letterpaper]{article}

\usepackage{mathrsfs}
\usepackage[utf8]{inputenc}
\usepackage{graphicx}
\usepackage[usenames, dvipsnames]{xcolor}
\usepackage[letterpaper,
            left=1in,
            right=1in,
            top=1in,
            bottom=1in]{geometry}
\usepackage{setspace}
\usepackage{textcomp}
\usepackage[symbol]{footmisc}
\usepackage{hyperref}
\usepackage{amssymb}
\usepackage{amsmath}
\usepackage{dsfont}
\usepackage[section]{placeins}
\usepackage{xcolor}
\usepackage[export]{adjustbox}
\usepackage{setspace}
\usepackage{natbib}
\usepackage{sectsty}
\usepackage{titlesec}
\usepackage[]{algorithm2e}
\usepackage{fancyhdr}
\usepackage{tabularx}
\usepackage{multirow}

\renewcommand{\abstractname}{ABSTRACT}
\renewenvironment{abstract}
 {\small
  \begin{center}
  \bfseries \abstractname\vspace{-.5em}\vspace{0pt}
  \end{center}
  \list{}{
    \setlength{\leftmargin}{0.5in}%
    \setlength{\rightmargin}{\leftmargin}%
  }%
  \item\relax}
 {\endlist}

\begin{document}

\titleformat*{\section}{\large\bfseries}
\titleformat*{\subsection}{\bfseries}
\titleformat*{\subsubsection}{\itshape}

\onehalfspacing
\begin{center}
\large{\textbf{Public transport challenges and technology-assisted accessibility for visually impaired elderly residents in urban environments}} 
\end{center}

\medskip

\begin{center}
\large{Jason Pan$^{\mathrm{a}}$, Ben Moews$^{\mathrm{a,b}}$}

\bigskip

\begin{small}
{$^{\mathrm{a}}$Business School, University of Edinburgh, 29 Buccleuch Pl, Edinburgh EH8 9JS, UK}\\
{$^{\mathrm{b}}$Centre for Statistics, University of Edinburgh, Peter Guthrie Tait Road, Edinburgh EH9 3FD, UK}
\end{small}

\end{center}

\bigskip

\doublespacing

\begin{abstract}
Independent navigation is central to social participation and health for vulnerable populations. While historic cities such as Edinburgh often feature well-established public transport systems, urban accessibility challenges remain and are exacerbated by complex landscapes, especially for groups with multiple vulnerabilities such as the visually impaired elderly. With limited research examining how real-time data feeds and artificial intelligence in this context, we address this gap through a mixed-methods approach. Our spatio-temporal analyses make use of statistical and machine learning techniques to investigate network coverage, service patterns, and density profiles through live-recorded data. This is combined with a qualitative thematic analysis of semi-structured interviews with the target group, as well as links to spatial cognition theory. The results demonstrate the highly centralised nature of the city's transport system, the significance of memory-based navigation, and the lack of travel information in usable formats. We also find that participants already use navigation technology to varying degrees and express a willingness to adopt artificial intelligence. Our findings highlight the importance of dynamic tools to meaningfully improve independent travel, as well as limitations due to the recurring problem of specific accessibility data, for example for facilities, often not being collected and stored.
\end{abstract}

\bigskip

\begin{footnotesize}



\end{footnotesize}

\bigskip


\section{Introduction}
\label{sec:introduction}

As ageing populations increase, the number of elderly individuals experiencing sight loss continues to grow, impacting their ability to confidently navigate urban environments \citep{Schulz2015, Wood2022}. In historic cities such as Edinburgh, public transport systems are crucial for mobility, but infrastructure obstructions and a lack of accessible real-time travel information discourage independent navigation for visually impaired people, increasing reliance on others and reducing quality of life \citep{Campisi2021}. With 10\% of Edinburgh's population being older adults, inclusive mobility solutions are highly relevant for the city's fastest-growing demographic \citep{EHSCP2022}. In this context, artificial intelligence (AI) is increasingly being explored for navigational utility, supporting independence and accessibility for age-related vision impairments \citep{MA2023101808, wong2025olderadults}. 

This study examines navigational challenges encountered by the visually impaired elderly using Edinburgh's public transport networks and assesses the potential of technological advances to address these barriers. We employ a mixed-methods approach combining quantitative analysis of transport data and qualitative insights from semi-structured interviews. Our aim is to understand whether, and how, these kinds solutions can address accessibility challenges and enhance navigation. We explore the question how existing infrastructure impacts independence, identify key day-to-day barriers, and explore attitudes and expectations regarding AI-assisted navigation.

In doing so, we provide multiple original contributions. While public transport for the visually impaired and elderly has been investigated separately, including for Edinburgh by \citet{Montarzino01012007} and \citet{GORMAN200315}, through qualitative research and surveys, respectively, the literature on the overlap with multiple vulnerabilities remains sparse. An exploration of the potential of current technology developments furthers these insights. Related quantitative work in the literature also generally relies on traditional descriptive statistics for insights, whereas the quantitative part of our work strongly focusses on spatial analysis through machine learning approaches to gain a different angle on a public transport network.


\section{Background}
\label{sec:lit-review}

\subsection{Barriers to inclusive urban accessibility} 

Accessibility is essential to urbanisation and characterised as a fundamental human right central to maintaining social equity and inclusion, especially for marginalised populations limited by mobility \citep{GILSOLA20181, UN-SUD}. This is reinforced as a fundamental right in the UN Convention on the Rights of Persons with Disabilities \citep{Nykiforuk2021}. As urban areas expand and ageing populations increase, insufficient accessibility continues to pose obstacles to achieving equity \citep{WANG2022103611}. 

These environments serve as sociocultural, economic, and technological centres, yet often fall short in that regard \citep{Pineda2024}. \citet{WANG2024103983}'s analysis of Hong Kong finds that, despite the city's transport-oriented design, location-based characteristics and overcentralisation results in uneven accessibility. With approximately 10\% of the world's population being seniors, this number is projected to double by 2050 \citep[see][]{UNWSR}, while more than half of the elderly live in cities \citep{Padeiro2021}. Recent studies describe barriers such as inconsistent and overcentralised transport services, and outdated or unmaintained pedestrian infrastructure \citet{ASIEDUAMPEM2024101935, EISENBERG2024104837}.

The number of people with vision impairments is expected to rise by about 60\% in 25 years, with an increased likelihood of age-related vision loss \citep{WernerWahl2013}. Vision loss varies in severity and can impact the ability to navigate cities independently, with everyday tasks relying on sight becoming daunting, placing visually impaired elderly people amongst the most marginalised groups in society \citep{Fraser2019}.

\subsection{Vision impairment and ethical considerations}

The experience of navigating urban environments is a cognitive task relying on sensory information for spatial awareness \citep{Meneghetti2021}. This becomes more complicated for ageing vision-impaired populations as they must rely on alternative sensory cues and adaptive strategies \citep{Schinazi2015}. \citet{Cimarolli2011} examine 364 individuals and reveal that over 60\% associate navigating in built environments with fear and insecurity. The inability to engage with surroundings leads to social isolation and reduced interpersonal relationships, which can exacerbate health issues \citep{Barnes2022}.   

Vision plays a key role, as humans navigate using two main strategies for spatial orientation, egocentric (self-centred) reliance on sensory cues and allocentric (environment-centred) navigation dependent on locations relative to oneself \citep[see, for example,][]{Ruggiero2021}, with age-related shifts towards the latter. For the visually impaired elderly, this dependence increases cognitive strain due to reduced memory capacity and slower cognitive processing, with assistive technology having been proposed to ease increased cognitive strain and sensory overload \citep{Mushin2024}.

Independence is central to well-being as a key aspect of the World Health Organization's Quality of Life (WHOQoL) framework \citep{WHOQoL}. For the elderly, especially those with additional challenges, maintaining independence can be challenging, impacting decisions such as domicile and level of environmental engagement \citep{Remillard2023}. The inconsistencies of cities introduce obstacles through heavy foot traffic, vehicle movement, and sensory overstimulation \citep{Vincent2024}.

\subsection{Public transport and navigating Edinburgh}

The elderly represent one of the UK's fastest-growing demographics at nearly 20\% of the population. Among them, around 2 million live with vision impairments \footnote{\url{https://www.nhs.uk/conditions/vision-loss}}\textsuperscript{,}\footnote{\url{https://www.rnib.org.uk/professionals/research-and-data}} As the capital of Scotland, Edinburgh is a major political, academic, and economic hub undergoing urban redevelopment and home to a growing elderly population estimated to make up nearly 10\% \citep{GaoEdiFestival, EHSCP2022}. The city's landscape and infrastructure present accessibility challenges due to physical limitations and complicated urban design \citep{HALDEN2002313, KAROU20141}. 

Edinburgh and the Lothians constitute approximately 16\% of Scotland's population, featuring the Edinburgh City Region or Greater Edinburgh Area \citep{Docherty2008}. While the mix of multi-era architecture offers cultural value and appeal, historic urban centres are prone to challenges for vulnerable populations \citep{su16062485, Miranda2020, COOKE20173003}. Efforts to revitalise historic cities typically rely on `top-down'  strategies, beginning with a central hub to prioritise zones for economic activity at the expense of outer regions \citep{Sun20200921, Ergenc15032024}.

The city's public transport network is governed by Transport for Edinburgh\footnote{\url{https://www.edinburghchamber.co.uk/membership/our-partners/transport-for-edinburgh}} (TfE), the UK's largest publicly operated bus company, and the City of Edinburgh Council's Public Transport Action Plan (PTAP) 2030 and City Mobility Plan (CMP) 2021-2030 both acknowledge areas for improvement. The CMP's `Vision Zero' initiative aims to improve accessibility and safety, with a target to eliminate traffic-related fatalities by 2050.\footnote{\url{https://www.edinburgh.gov.uk/city-mobility-plan-1}} Given the city's bus-heavy transport network, a comprehensive understanding of how visually impaired elderly people experience the city is essential for these projects.

\subsection{Artificial intelligence in urban navigation}

AI methods are experiencing rapid growth with the potential to transform operations across sectors like healthcare and transport \citep{DWIVEDI2023122579}. In this context, smart cities apply modern technology to enhance quality of life, improve efficiency in urban operations, and maintain sustainable growth \citep[see][]{Batty20121009}, with related work showing how they can enhance urban mobility through real-time information and sensor-based infrastructure \citep{ELASSY2024100252, KOLOTOUCHKINA2022103613}.

Real-time transport data can enhance navigation aids by providing up-to-date information about vehicle arrivals, route changes, and surrounding areas. Access to live information may reduce uncertainty when travelling and support more confident decision-making. For the visually impaired elderly, this can lessen reliance on others and allow for greater autonomy in everyday navigation \citep{ABIDI2024e31825}.

Aligning with Edinburgh's smart city initiative, the City of Edinburgh Council has trialled accessible technologies to assist visually impaired users by activating pedestrian crossings via Bluetooth through a mobile app \citep{IETneatebox}. Here, a major drawback is the need to retrofit pedestrian crossings with specific hardware. Similar trials in Malaysia and Australia are observed by \citet{Deverell03042022} but raise further concerns about user training requirements, a lack of focus on existing personal navigation aids, and the prerequisite that users own and confidently use smartphones.

Despite advancements demonstrating the utility of AI in healthcare \citep[see, for example,][]{MA2023101808}, ethical considerations must be addressed for handling sensitive information. Aside from robust security measures and the prevention of data misuse, data collection and usage should be transparent and the associated benefits and risks made clear to patients \citep{Dankwa-Mullan2024}. Safeguards around consent and algorithmic fairness are essential to ensure that AI-driven solutions are accessible, equitable, and trustworthy. 

\subsection{Existing tools for the visually impaired elderly}

Conventional navigation aids such as the well-established white cane have long been essential tools \citep{rasoulikahakiCane2023}. When examining strengths and weaknesses, \citet{ABIDI2024e31825} note that the latter are not as effective at detecting non-ground-level features of urban landscapes, such as dangers above, in front, and parallel to one's head. Guide dogs provide not only mobility but also emotional support, but drawbacks include the inability to interpret signs or traffic signals and overstimulation in busy areas \citep{Whitmarsh01012005}. While traditional aids have been effective, they fall short in adapting to fast-paced situations requiring real-time information and dynamic alerts \citep{prandiAccessibleWayfindingNavigation2023}. 

AI research for accessibility and transport is not a novel concept; wearable technology and real-time devices have been adapted to complement existing systems, with smart sticks being one example. \citet{SmartStickResearch} develop and trial such a tool with ten elderly participants, featuring obstacle detection, GPS functionality, and mobile app integration. Device personalisation in their work is limited to changing sensor sensitivity levels, with haptic or audible navigation feedback, as a cost-effective and user-friendly alternative to guide dogs in developing countries. However, limitations include the inability to detect pits in roads and elevation changes, which are key challenges to accessibility in Edinburgh's landscape. 

Research on real-time AI solutions includes \citet{kulkarniBusStopCVRealtimeAI2023}, who use use computer vision to label accessibility features of bus stops, identifying whether they are enclosed by shelters or open, and whether they have timetables, signage, or seating. However, the tool relies on a crowd-sourcing network of users to contribute updates and verify live information, similar to how Google Maps asks its users questions to update live data and provide real-time crowd tracking \citep{PARADY2023344}. 


\section{Methodology}
\label{sec:methodology}

\subsection{Research approach and data collection}
\label{sec:data-collection}

We adopt a mixed-methods approach integrating quantitative analysis of transport network structure with qualitative insights from participants. The quantitative analysis identifies spatio-temporal service patterns, while qualitative interviews are used to interpret how these patterns are experienced by visually impaired users, reflecting related works \citep{WONG2018300, ABERLE2025104140}. Interview themes concerning barriers, uncertainty, and information needs are contextualised using the identified spatial structure.

Quantitative data collection spans from 2$^\text{nd}$ March 2025, 20:00 to 10$^\text{th}$ March 2025, 20:00 in five-minute intervals through the TfE Open Data API\footnote{\url{https://tfe-opendata.readme.io}}. Using real-time vehicle location data, rather than static timetable information, allows observation of actual service patterns and variability as experienced by users. This is particularly relevant when assessing accessibility for vulnerable groups, as deviations from scheduled services, temporary disruptions, and uneven service frequency can considerably affect travel reliability.

We delete rows with null values indicating inactive vehicles, verified via null travel direction values and coordinates for the Lothian Bus depot, the tram depot at Gogar, and the maintenance centre at Seafield, as well as inactive routes through missing next stops and destination values. We interpret the overall network structure using the spatial measures introduced in Section~\ref{sec:data-analysis} in relation to the residential distribution of elderly residents in Edinburgh to assess areas with accessibility needs \citep{EHSCP2022}.

By necessity, our work focusses on accessibility using spatial and temporal data, following the literature facing the same data availability challenge and where geospatial distributions are commonly the focus of investigations. As such, we define (spatio-temporal) accessibility for the purpose of this study as reachability in terms of distance and service frequency. We discuss the implications and limitations of this, as well as the distinction from information and facility accessibility, in Sections~\ref{sec:limitations} and~\ref{sec:discussion}.

Our qualitative analysis uses semi-structured phone interviews as a non-intrusive method, which reduces the stigma, discomfort, and potential bias of formal research settings \citep{Emara2025, Akram2021}. Participant recruitment follows a purposive sampling approach through our contact with the Royal National Institute of Blind People (RNIB) Scotland and North East Sensory Services, after ethics approval \citep{AHMED2024100662, Stratton2024}. Only individuals contacted through these organisations are featured in the study. 

One participant is aged 60--70 and the remaining 70--80; all are registered as fully blind under UK and Scottish certification standards, with one participant developing vision impairment in later life and the remaining with congenital blindness. They reside within the Edinburgh region including both central and peripheral areas. All participants are informed of the study through a verbal explanation of the participant information sheet and consent form, and verbal consent was obtained and recorded separately. Interviews are transcribed and cleaned manually using verbatim-style transcription \citep[see][]{McMullin2023}.

Interviews target public transport experience, with questions adapted from multiple studies, including \citet{wong2025olderadults} and \citet{Shehri2022}. Our research involves vulnerable participants and Level 3 ethics approval was granted by the university's corresponding ethics committee. More precise demographics or locations are not reported to protect participant anonymity, in line with the University of Edinburgh's ethics and data protection regulations. Participants are informed of the study's purpose and any traceable personal information is redacted from transcripts.

\subsection{Quantitative data analysis}
\label{sec:data-analysis}

Introduced by \citet{Clark1954}, we apply nearest neighbour analysis (NNA) to examine whether the distribution trend of points is due to clustering, randomness, or dispersion calculated by
\begin{equation}
    R = \frac{\bar{r}_{A}}{\bar{r}_E} \ \mathrm{s.t.} \ \bar{r}_{A} = \frac{\sum_{n = 1}^{N} r}{N}, \ \bar{r}_E = \frac{1}{2\sqrt{\rho}},
\label{eq:nearest_neighbour_analysis}
\end{equation}
where $R$ is the Nearest Neighbour Index (NNI), $\bar{r}_A$ is the average distance to the nearest stop, $\bar{r}_E$ is the expected average distance under a random distribution, and $r$ represents each nearest-neighbour distance. Lastly, $\rho$ is the point density, calculated as the number of stops $N$ over the area of interest $A$,
\begin{equation}
    \rho = \frac{N}{A}.
\label{eq:point_density}
\end{equation}
We compute the nearest-neighbour distances between all stop coordinates to calculate the NNI. Additionally, a $Z$-test assesses the strength of $R$, using the formulation proposed by \citet{Clark1954},
\begin{equation}
    Z = \frac{\bar{r}_A-\bar{r}_E}{\sigma_{\bar{r}_E}} \ \mathrm{s.t.} \ \sigma_{\bar{r}_E} = \frac{0.26136}{\sqrt{N\rho}}.
\end{equation}
We also use Ripley's $G$ and $F$ functions to explore the spatial non-uniformity \citep{Ripley1976, Ripley1977}. The former calculates, for different distance thresholds $\tau$, the cumulative distribution of nearest-neighbour distances $\delta_\mathrm{n}$,
\begin{equation}
\hat{G}(\tau) = \frac{1}{n} \sum_{i=1}^n \mathds{1} \{ \delta_\mathrm{n}(D_i) \leq \tau \} \ \mathrm{s.t.} \ \delta_\mathrm{n}(D_i) = \min_{i \neq j} || D_i - D_j ||,
\label{eq:ripley_g}
\end{equation}
for a distance matrix $D$, which can then be compared to a homogeneous Poisson process $G_p(\tau)$ so that $\lambda$ and $\hat{G}{\tau} > G_p(\tau)$ indicating the clustering of data points. Similarly, Ripley's $F$ function is defined as
\begin{equation}
\hat{F}(\tau) = \frac{1}{m} \sum_{i=1}^m \mathds{1} \{ \delta_\mathrm{e}(u_i, D) \leq \tau \} \ \mathrm{s.t.} \ \delta_\mathrm{e}(u, D) = \min \{ || u - D_i || : D_i \in D \},
\label{eq:ripley_f}
\end{equation}
using the empty space distance $\delta_\mathrm{e}$ for a random location $u \in \mathbb{R}^2$ and a grid $\{ u_1, u_2, \dots, u_m \}$, and an equivalent comparison to a homogeneous Poisson process. In contrast to Ripley's $G$ function, non-uniform clustering is indicated for $\hat{F}(\tau) < F_p(\tau)$. For an in-depth overview, see \citet{Baddeley2015}.

Kernel density estimation (KDE) is a nonparametric method that is useful when assumptions about the underlying probability distribution cannot easily be made \citep{Rosenblatt1956, Parzen1962}. We use vehicle data to analyse city-wide usage, with coordinates assigned to their nearest stop and the cumulative number of positions representing usage intensity. We make use of a bivariate kernel function to estimate the density of vehicle stop activity $\hat{f}(x,y)$ at a given point $(x,y)$, 
\begin{equation}
    \hat{f}(x,y) = \frac{1}{nh_xh_y} \sum_{i=1}^n K \left (\frac{x_i-x}{h_x},\frac{y_i-y}{h_y} \right ),
\label{eq:kernel_density_estimation}
\end{equation}
where $K$ is a kernel function \citep[typically Gaussian, see][]{Silverman1986}), $h_x$ and $h_y$ are bandwidth parameters for the respective dimensions, and $(x_i, y_i)$ are the $n$ vehicle coordinates \citep{YU201480}.
    
Lastly, we apply $k$-means clustering to stops and incorporate the vehicle–stop assignment data used in KDE \citep{MacQueen1967, Lloyd1982}. We section stops into $K$ clusters through similarity grouping \citep[see][]{CHEN2024110165}, which minimises the distances between data point and their randomly assigned centroid,
\begin{equation}
    \min \sum_{k=1}^K \sum_{x_i \in c_k} \left\| x_i - \mu_k \right\|^2,
\label{eq:similarity_grouping}
\end{equation}
where $x_i$ is a transport stop, $\mu_k$ is the centroid of cluster $k \in K$, and $c_k$ represents cluster $k$'s assigned stops. We use three features; latitude and longitude of stops and vehicle stop frequency. These features are standardised using $Z$-score normalisation to remove the mean and scale to unit variance,
\begin{equation}
    x' = \frac{x - \bar{x}}{\sigma}.
\end{equation}
Here, $x'$ is the original feature vector, and $\bar{x}$ and $\sigma$ are the mean and standard deviation to ensure equal weighting. Incorporating frequency in addition to two-dimensional coordinates allows the algorithm to consider physical stop distribution and service intensity, distinguishing between high and low-frequency stops that are geographically close but have different patterns \citep{CHEN2024110165}. 

\subsection{Qualitative data analysis}
\label{sec:qualitative}

We implement a hybrid thematic analysis \citep[see][]{Proudfoot2023}, with the deductive approach identifying themes and codes rooted in existing literature and a priori assumptions about AI and accessibility for the visually impaired elderly. Deductive themes serve as a framework for a comparison with inductive themes and codes that emerge from participants' verbatim, ensuring that findings are shaped by the data rather than preconceptions \citep{Fereday2006}.
\begin{table*}[!h]
\caption{Overview of the relationship between deductive assumptions and inductively developed main and sub-themes identified through the second-cycle thematic analysis.}
\label{tab:thematic_analysis}
\begin{center}
\begin{small}
\begin{tabularx}{\textwidth}{|X|X|c|}
\hline
\textbf{Deductive Assumptions} & \textbf{Sub-Themes} & \textbf{Main Themes} \\
\hline
\hline
Many visually impaired elderly rely on sighted assistance for navigation. & Dependence on Support & \multirow{3}{*}{Navigational Dependence} \\
\cline{1-2}
Infrequent, unreliable, or complex services cause travel difficulties. & Barriers to Independence & \\
\hline
Current urban infrastructure is inadequate for accessibility. & Urban Accessibility & \multirow{3}{*}{Transport Accessibility Challenges} \\
\cline{1-2}
Pedestrianisation and transport changes disrupt familiar routes and discourage outdoor activity. & Physical and Environmental Barriers & \\
\cline{1-2}
Lack of staff support and clear information make it difficult to use public transport independently in Edinburgh. & Limited Resources and Support & \\
\hline
Visually impaired elderly are generally unfamiliar with new technologies. & 
Technological Familiarity & \multirow{3}{*}{Technology Readiness and Confidence} \\
\cline{1-2}
Most have limited or no prior exposure to AI-based tools. & Adoption Readiness & \\
\hline
Increased sensory dependence and established practices creates scepticism towards technology. & Memory-Based Reliance & \multirow{3}{*}{Cognitive and Sensory Demands} \\
\cline{1-2}
Fear of judgement and dependence discourages visually impaired elderly from asking for assistance & Psychological Impact & \\
\hline
Visually impaired elderly fear that technology could reduce meaningful human interaction. & Reduced Human Interaction & \multirow{3}{*}{Perceptions of Trust, Privacy, and Responsibility in AI} \\
\cline{1-2}
Many fear how their personal data may be used. & Attitudes Towards Data Privacy & \\
\hline
There is low understanding of AI and reluctance towards adoption. & General Attitudes Towards AI & \multirow{3}{*}{Perceptions of AI-Assisted Navigation} \\
\cline{1-2}
Participants have clear expectations for future accessible technology. & Expectations and Benefits of AI Use & \\
\hline
\end{tabularx}
\end{small}
\end{center}
\end{table*}

We follow the two-cycle thematic analysis framework by \citet{braun-clarke2017}. Interview transcripts are read repeatedly to identify segments relating to barriers, independence, technology use, trust, and sensory or cognitive demands. These segments are assigned descriptive first-cycle codes, adapted from \citet{Thomas2008}, which are compared across participants to identify similarities and differences. Related codes are grouped into broader second-cycle themes as shown in Table~\ref{tab:thematic_analysis}. Given the sample size, the aim is not to establish statistical representativeness but to provide exploratory, lived-experience-based insights.

\subsection{Methodological limitations}
\label{sec:limitations}

The TfE API and data availability present several challenges. Stops and live vehicle locations are available but lack attributes such as vehicle delays, stop congestion or service disruptions. This restricts the ability to analyse live accessibility barriers; a more granular version was previously offered. Live data became unavailable in February 2025 due to maintenance and has, at the time of writing, not been restored, impacting the ability of the research community to fully explore the transportation system.

Moreover, a Freedom of Information request under the Freedom of Information (Scotland) Act 2002\footnote{\url{https://www.legislation.gov.uk/asp/2002/13/contents}} to the City of Edinburgh Council prompted the clarification that accessibility data is not stored and, as such, remains unavailable retroactively for research. Consequently, the level of inclusivity cannot be assessed, further complicating the analysis of vulnerable groups; it is uncertain whether the council consistently achieves adequate service accessibility. At the time of writing, the TfE API has also stopped providing any accessibility information, real-time or otherwise, which exacerbates this challenge.

For our quantitative data collection, although disruptions reflect realistic variability, the observation period clashes with the northbound North Bridge closure, introducing some abnormal service patterns. Recruiting participants is also challenging due to the sensitive nature of interacting with vulnerable persons willing to participate \citep{Ellard2015}.


\section{Findings}
\label{sec:findings}   

\subsection{Edinburgh's public transport system}
\label{sec:quant-findings}

We first analyse real-time vehicle locations, with 613,538 recorded data points covering 3,142 unique stops and 103 individual services. The most frequently used services are presented in Table~\ref{tab:common_services}. 

\begin{table*}[!h]
\caption{Most frequent transport services in Edinburgh with the highest recorded vehicle location instances, indicating the most active routes during the collection period. T50 represents the Edinburgh Tram service.}
\label{tab:common_services}
\begin{center}
\begin{small}
\begin{tabular}{|l||r|r|r|r|r|r|r|r|r|r|}
\hline
Service & 26 & 30 & 44 & 3 & 16 & T50 & 21 & 25 & 11 & 31 \\
\hline
Count & 27995 & 23675 & 22684 & 21853 & 20742 & 19433 & 17213 & 17198 & 17133 & 17092 \\
\hline
\end{tabular}
\end{small}
\end{center}
\end{table*}

Top entries indicate high-demand journeys connecting the Old and New Town via Princes Street, while 8.4\% of buses are marked as `Not in Service'. We remove entries with null or N/A values for service names and conduct an NNA on stops, with the NNI calculated via Equation~\ref{eq:nearest_neighbour_analysis}. As established by \citet{Clark1954}, an $R = 0$ indicates that points are all in the same location, whereas $R \approx 1$, $R < 1$ and $R > 1$ imply randomisation, clustering and dispersion, respectively. The results are summarised in Table~\ref{tab:nna_results}. 

\begin{table*}[!h]
\caption{Summary of the Nearest Neighbour Analysis (NNA) results for the stops data described in Section~\ref{sec:data-collection}, measuring the spatial distribution of public transportation system stops in Edinburgh.}
\label{tab:nna_results}
\begin{center}
\begin{small}
\begin{tabular}{|l||r|}
\hline
Observed Average Nearest Neighbour Distance ($\bar{r}_A$) & 110.225 (metres) \\
\hline
Expected Average Nearest Neighbour Distance ($\bar{r}_E$) & 642.191 (metres) \\
\hline
Nearest Neighbour Index ($R$) & 0.172 \\
\hline
$z$-score & -97.691 \\
\hline
$p$-value & $3.48 \times 10^{-2075}$ \\
\hline
\end{tabular}
\end{small}
\end{center}
\end{table*}

\begin{figure*}[!htb]
\centering
\includegraphics[width=\textwidth]{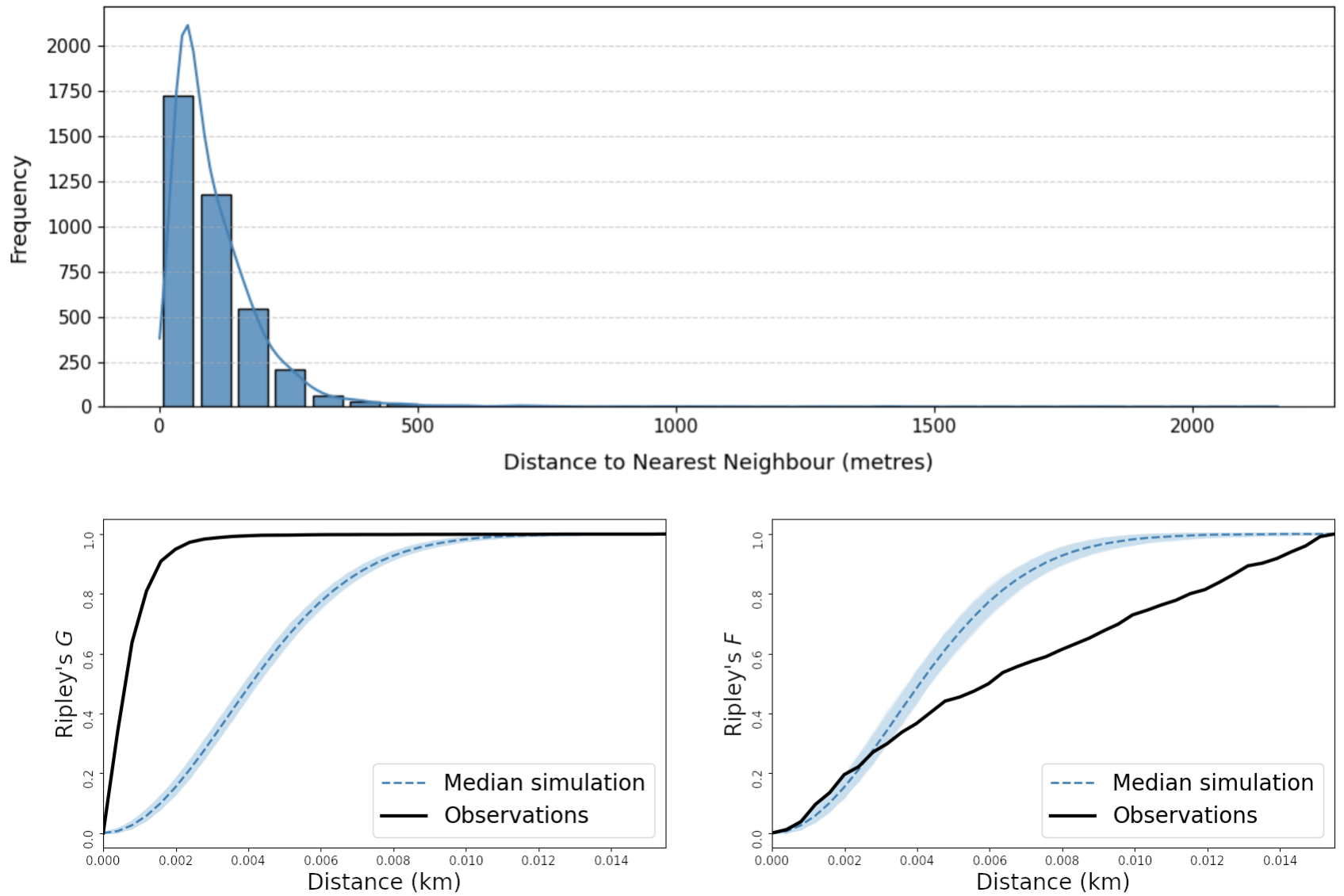}
\caption{Distance measurements across the Edinburgh transport system. The upper panel shows the frequency distribution of nearest-neighbour distances between all stops (bus and tram), with the overlaid smoothed line representing a kernel density estimate. The lower left and right panel show Ripley's $G$ and $F$ functions for distances, respectively, as solid lines, with the dashed line depicting the median of Poisson simulations and their 95\% confidence intervals as shaded regions.}
\label{fig:dist_nna}
\end{figure*}

The results indicate strong and statistically significant clustering in stop locations. We evaluate the significance through $z$-scores, for a null hypothesis of the distribution of $\bar{r}_A$ being approximately normal and a significance threshold of $\alpha = 0.01$, which we reject. The distribution of nearest-neighbour distances is shown in Figure~\ref{fig:dist_nna}, featuring a strongly right-skewed distribution.

Ripley's $G$ function, as the proportion of samples for which the nearest other data point is within a threshold distance, rises much faster to near-complete coverage than the homogeneous simulation, providing a measure of ``clumpiness'' in the spatial distribution. Similarly, the almost linear increase of Ripley's $F$ function indicates large empty spaces in the spatial distribution, with distances from random points located in such gaps leading to a slower rise compared to the homogeneous simulation.

\subsection{Insights into stop density patterns}
\label{subsec:kde}

\begin{figure*}[!htb]
\centering
\includegraphics[width=\textwidth]{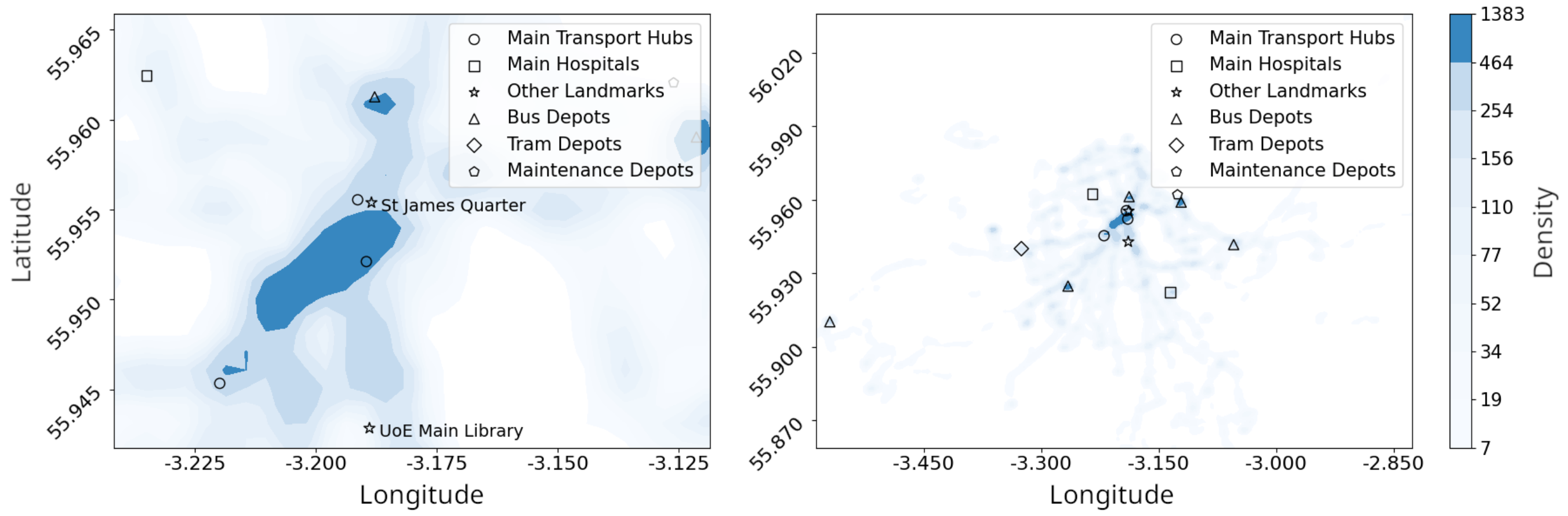}
\caption{Kernel density estimate for transport stop density in Edinburgh. The left-hand panel shows the Greater Edinburgh Area, while the right-hand panel depicts a zoomed-in view of the City Centre. Key locations for transport and public life are marked separately.}
\label{fig:kde_comparison}
\end{figure*}

We then apply KDE to the combined data on stops and vehicle locations. The resulting values correspond to the cumulative number of vehicles over time at their closest assigned stops, overlaid with key locations such as transport hubs, landmarks, and hospitals, in Figure~\ref{fig:kde_comparison}.

The highest-density values above 1,000 fall around major transport hubs such as Waverley Station, Haymarket, and Princes Street. Similarly, regions such as Leith, North Bridge, Abbeyhill, and the West End exhibit moderate values ranging from 110 to 464, and lower densities (less than 77) in peripheral areas such as Balerno, Wester Hailes, and Dalkeith corresponding to predominantly residential areas, including neighbourhoods with higher proportions of elderly residents \citep{EHSCP2022}. This spatial pattern provides context for the interview findings regarding unevenness in access discussed in Section~\ref{sec:qual-findings}, particularly participants' concerns about indirect journeys, reliance on central interchange stops, and difficulties accessing key destinations without assistance.

We run preliminary experiments for different smoothing bandwidths $h \in [0.1, 0.3, 0.5, 0.8, 1.0]$ to evaluate their influence on the density interpretation, as lower $h$ values produce more detailed but potentially noisy maps. We find that $h > 0.5$ leads to excessive smoothing that misses discrepancies between the central and suburban areas, while $h < 0.3$ introduces excessive granularity, making the results harder to interpret. As a result, we set $h=0.3$ for the granularity of suburban areas and central stop-dense areas.

\subsection{Cluster analysis of stops and usage} 
\label{subsec:kmeans}

\begin{figure*}[!htb]
\centering
\includegraphics[width=\textwidth]{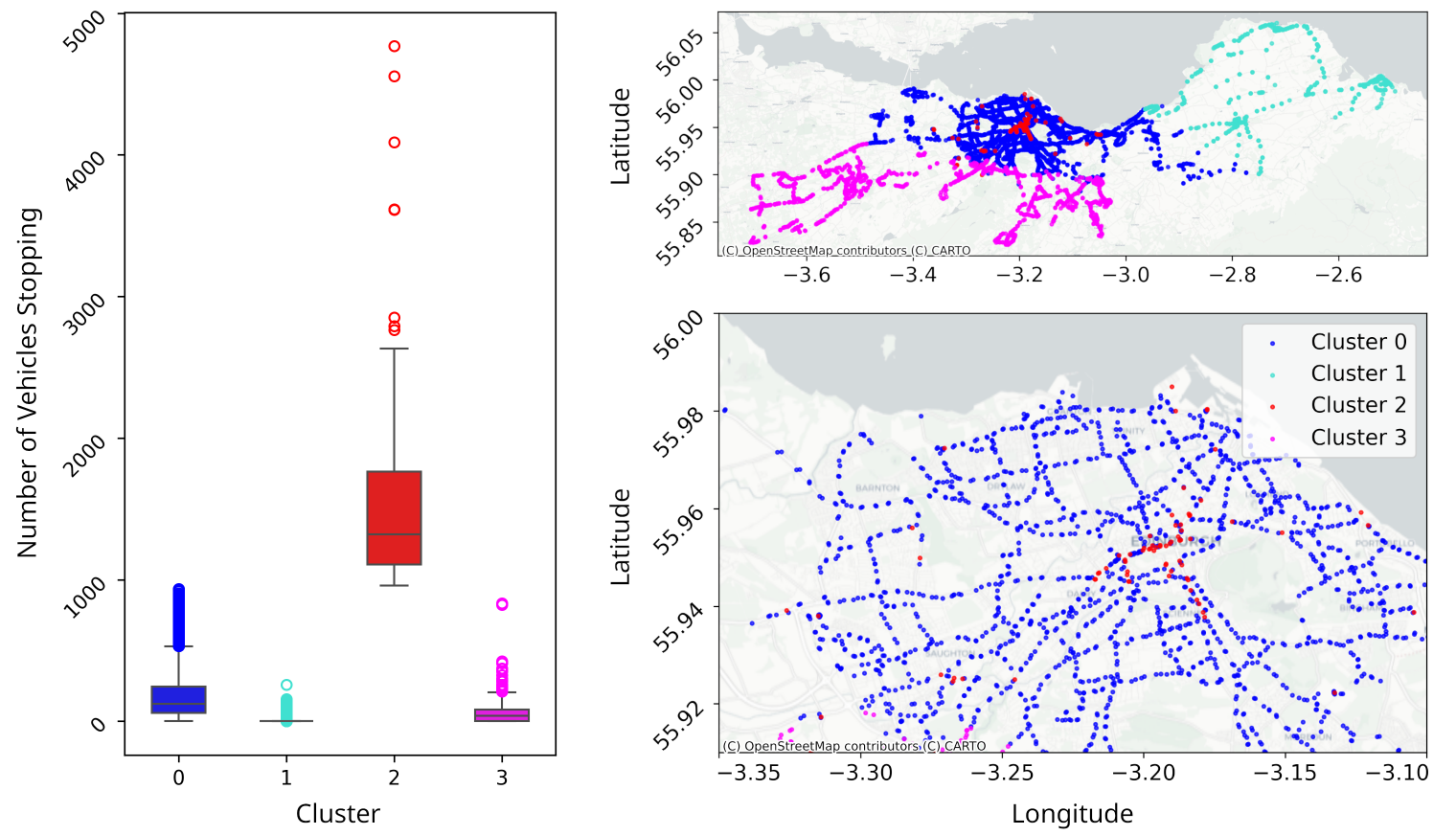}
\caption{Cluster analysis results for geographical locations and usage frequency (activity) of stops. The left-hand panel shows a boxplot representing the distribution of activity within each cluster, measured by the number of vehicles frequenting each stop. An extreme outlier in Cluster 2 with a value of 8,708 vehicles is excluded from the plot, corresponding to the central area where many key interchange stops are located. The upper right-hand panel depicts the clustering of transport stops in the Greater Edinburgh Area based on both location and activity, while the lower right-hand panel shows the same for the City Centre.}
\label{fig:cluster_analysis}
\end{figure*}

Finally, we employ $k$-means clustering to identify stop accumulations based on coordinates and usage frequency. The choice of $k = 4$ is the result of iterative cluster stability tests, showing that $k > 4$ results in oversegmentation, whereas $k < 4$ leads to overly broad groupings that fail to accurately differentiate usage frequency levels by location. The resulting Cluster 0 features 2,198 stops, Cluster 1 422 stops, Cluster 2 86 stops, and Cluster 3 1,094 stops. We analyse their characteristics through boxplots and geospatial scatter plots for both the Greater Edinburgh Area and the City Centre, as depicted in Figure~\ref{fig:cluster_analysis}. 

While Cluster 0 contains the greatest number of stops, Cluster 2 exhibits far higher service frequency; a higher stop count does not correspond to higher transport activity. Cluster 3 represents moderate-frequency stops, while Cluster 1 exhibits the lowest activity, and outliers for very-high-frequency stops are especially notable in Cluster 2. Incorporating vehicle counts provides insights into stop usage frequency and how activity levels are physically distributed compared to clustering solely based on geographic proximity.

\begin{figure*}[!htb]
\centering
\includegraphics[width=\textwidth]{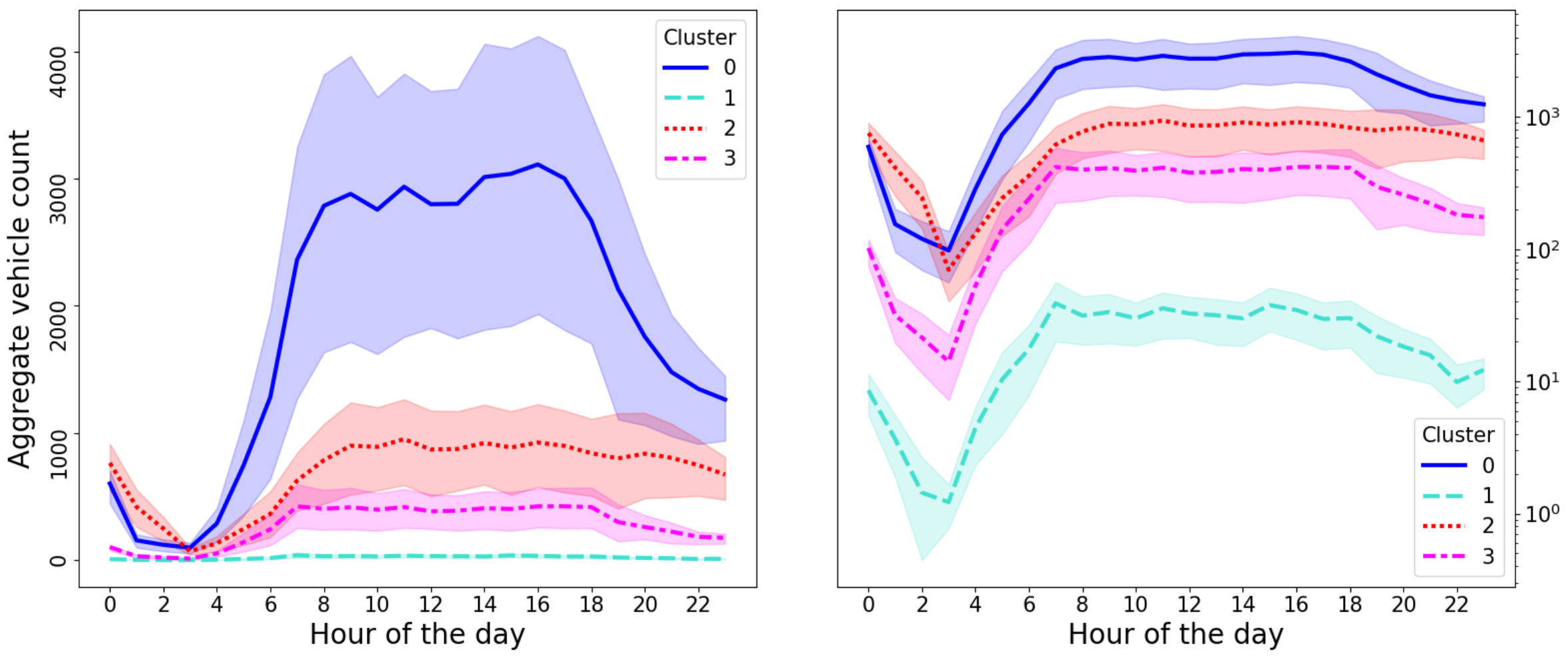}
\caption{Aggregate vehicle counts per cluster. The left-hand figure shows the total of vehicles visiting stops per hour of the day for each cluster, with line types also described in the legends, and with dark lines indicating the arithmetic average across the eight observed days and lightly-shaded enveloped depicting the 95\% confidence intervals due to inter-day variation. The right-hand figure depicts the same, but on a logarithmic y-axis to enable shape comparability between clusters with different totals.}
\label{fig:temporal_visualisation}
\end{figure*}

Lastly, we also visualise the stop frequency evolution across an averaged 24-hour period, in Figure~\ref{fig:temporal_visualisation}. Clusters are neatly split between stop volumes, with correspondingly increasing confidence intervals for higher numbers. For a comparison of evolution shapes, the right-hand panel employs a logarithmic scale, showing an expected dip around 3am in the night for minimal service availability. Overall, clusters evolve similarly, with the exception of Cluster 2 providing a higher total count than Cluster 1 between midnight at 2am, despite a smaller total number of stops, due to containing many highly-active stops in the inner city.

\subsection{Qualitative findings}
\label{sec:qual-findings}

Our analysis of semi-structured interviews uses a two-stage coding approach, following a systematic framework for thematic analysis \citep[see][]{braun-clarke2017}. Our findings help to explain how spatial patterns are experienced in practice, especially where centralised service activity, peripheral thinning in relevant residential areas, and crowded hubs shape travel decisions. We note differences in terms of technology familiarity and residential context, but avoid stronger subgroup comparisons due to the small sample size.

\textit{Navigational Dependence} emerges as a theme, with mixed perceptions, while sub-themes include \textit{Dependence on Support} and \textit{Barriers to Independence} for challenges in navigating difficult environments such as central Edinburgh, national rail stations, and transport hubs. Participant 1 (P1) affirms confidence before explaining, ``[...] when you're on the bus or train, you tend to be on your own, and a lot of people prefer to have their independence rather than ask a member of the public. So if I could do it on my own, I prefer to try to do it on my own.'' However, other participants expressed a reliance on physical assistance. 

\textit{Transport Accessibility Challenges}, like other themes, are interrelated with other challenges. We identify \textit{Urban Accessibility}, \textit{Physical and Environmental Barriers} and \textit{Limited Resources and Support} as sub-themes, aligning with findings from \citet{Chemnad2024}. Participants describe difficulties due to obstructions, a lack of landmarks, and construction, and excerpts from interviews are listed in Table~\ref{tab:interview_excerpts}. 

\begin{table*}[!h]
\caption{Thematic analysis of Accessibility Barriers as a theme during semi-structured interviews with elderly and visually impaired participants in the public transport system of Edinburgh.}
\label{tab:interview_excerpts}
\begin{center}
\begin{small}
\begin{tabularx}{\textwidth}{|X|X|c|}
\hline
References & Sub-Themes (First-Cycle Codes) & Themes (Second-Cycle Codes) \\
\hline
\hline
``[...] if it's a busy stop, in Edinburgh especially, it shows you the next seven eight buses that's on that screen [...] as the time [passes], they scroll up to the next one. I just really want to know my one. I'm not really interested in the one 10 minutes or 15 minutes either side of my one.'' -- P1 & Limited Resources and Support & \multirow{3}{*}{Transport Accessibility Challenges} \\
\cline{1-2}
``Well, I've got a good memory and it was [all] described to me so I can follow it quite well ... the fact that I had sight [previously] ... I could visualise ... But if someone who had been blind from birth had great difficulty visualising that.'' -- P2 & Urban Accessibility & \\
\cline{1-2}
``It's hard to get a picture in your mind of the layout of something, you know it might be a `Y' shape or a `T' shape or something and if you can't see and people don't know how to explain it [well], sometimes it's difficult to visualise [it] ...'' -- P3 & Physical and Environmental Barriers & \\
\hline
\end{tabularx}
\end{small}
\end{center}
\end{table*}

Participants recount challenging experiences with infrastructure, noting that digital interfaces at some stops lacked accessibility features to retrieve information. P1 states ``I want that information without having to learn how to use a device to get it. Now I know that information is there, but it's maybe not there in a format that would be suitable for blind [elderly] people to use.'' Additionally, P1 talks about information overload with existing transport infrastructure stating that ``I just really want to know my [bus]". This is particularly relevant to the high-activity clusters shown in Section~\ref{subsec:kmeans}, where multiple services converge.

Lack of staff is also discussed; P2 explains ``[...] they used to have the driver and conductor or conductress, and when you went on the bus the conductor or conductress [help] you get to a seat. But now they're all one man, just a driver, so [it's] very difficult [to] find the seat without somebody helping.'' They describe that inadequate services, frequent delays and shifting routes are common and poorly communicated.

For \textit{Technology Readiness and Confidence}, pre-analysis assumptions include potential technology unfamiliarity, but participants describe varying usage degrees, with identified sub-themes \textit{Technological Familiarity}, and \textit{Adoption Readiness}. P1 states ``I've used various types of technology ... I've used standalone GPS apps where it tells you your location and guides [you] from A to B ... I've got one called BlindSquare, which is an app on a phone and that allows me to put in a postcode or an address. But I can walk the route without leaving my house ... Yeah, it definitely does help.'' P2 exhibits openness to adoption, saying ``I suppose though over the years I've taken on various challenges, so if I had to I probably would.'' Similarly, P1 reaffirms a willingness to integrate AI as ``I think I would still rely on my own sensory abilities to do things that the technology [...] I would say the technology would be a backup.''

Beyond physical and digital barriers, participants discuss \textit{Cognitive and Sensory Demands}, including inaccessible information presentation increasing sensory interdependence \citep{Schinazi2015}. \textit{Memory-Based Reliance} and \textit{Psychological Impact} are key sub-themes; P1 expresses frustration with tracking apps as `top-heavy', saying they ``[...] fail with [the] uniformity [of] how information is presented.'' P2 highlights memory-based strategies on buses, using mental maps of topography and street layouts in high-traffic areas, describing trial and error with heightened alertness. All emphasise auditory information and sensory interdependence, with P1 and P3 stating ``[...] you're constantly listening ... You're using your ears more'' and ``[...] it  would depend how [the] information comes'', respectively.

Regarding \textit{Perceptions of Trust, Privacy, and Responsibility in AI}, participants employ traditional navigation aids such as canes and guide dogs. Sub-themes of ethical considerations are \textit{Attitudes Towards Data Privacy}, and \textit{Reduced Human-Interaction}. P2 states that while they have ``nothing to hide'', they recognise that some are ``very sceptical about it''. Participants express concerns about job displacement and interactions with sighted guides, and place data security expectations onto corporations, with P3 insisting ``[...] companies have a responsibility to deal with it [the data] responsibly. It's not my responsibility.''

Lastly, discussions cover practical applications under the theme of \textit{Perceptions of AI-Assisted Navigation}; participants demonstrate varying attitudes, with some viewing AI as a tool to push beyond the limits of their senses. P1 describes a hypothetical tool that provides real-time adaptive feedback to guide them through disruptions and suggest alternatives, stating ``[...] the AI knows that 200 yards from where it's broken down there's a bus that goes to my house. It's doing the thinking for me'', adding that ``[technology] revolutionises your ability to do things that you couldn't have done before.'' 

Others discuss enhancing spatial awareness, with P2 noting``If there was something you could get, like an audio map ... telling you [where] places were on a street or something like that, it could be quite helpful.'' This aligns with the sub-themes \textit{General Attitudes Towards AI}, and \textit{Expectations and Benefits of AI Use}, which all describe to varying extents. While P3 displays indifference, others note the importance of usability technological learning curves, as P2 expresses difficulty learning to handle touchscreen devices.


\section{Discussion}
\label{sec:discussion}

The first question posed in the introduction deals with the way Edinburgh's existing infrastructure impacts independent navigation, with our findings confirming a high degree of centralisation and notable contrasts between clustered and dispersed points. These results align with \citet{Miranda2020} on overcentralisation trends of urban redevelopment approaches in historic cities, with $k$-NN patterns that are consistent with urbanisation strategies where central interconnectivity efforts are prioritised over suburban expansion \citep{COOKE20173003, Sun20200921}. 

Our findings can also be interpreted through complementary perspectives in transport geography, as r integrating network science and transport accessibility argues that network structure alone is insufficient for understanding public transport performance and that service intensity, waiting times, transfers, and travel impedance are central to accessibility outcomes \citep{LUO2019102505}. Observed clusters and uneven service intensity similarly indicate that accessibility depends on how service concentration shapes movement across the city, aligning with travel behaviour theory emphasising that mobility choices are also shaped by perceptions, attitudes, preferences, and repeated behaviours \citep{VanAcker01032010}.

Our KDE analysis on the thinned-out peripheral coverage agrees with participants expressing unfavourable perceptions of accessibility. Specifically, P2 and P3 assert that visually impaired elderly people often avoid public transport due to a lack of services connecting outlying residential areas to hospitals and other key locations without having to switch services in the centre. As the majority of Edinburgh's elderly population resides in the outer North West and South West regions, this leads to long walking distances and potential delays \citep{EHSCP2022}. This links back to the theory outlined above, as peripheral thinning can increase the perceived and practical burden of reaching essential destinations. Taken together, the quantitative and qualitative findings show that accessibility barriers arise from both the spatial organisation of the network service and the lived difficulty of interpreting and using that network under conditions of uncertainty.

Our second question addresses key day-to-day barriers such as the lack of real-time information. Participants mention the lack of audio or Braille formats at stops for updates, and P1 describes overwhelming experiences with poor app customisability, aligning with the work by \citet{Remillard2023}. Overreliance on memory-based navigation strategies also appears, with P2 discussing mental cues, stating ``[...] but [on] the buses, when you ask the drivers, sometimes they remember to tell you but other times they [don't]. But it was handy [if] I was going to Waverley station because when the bus went down [onto] Princes Street, [when] you turned left [onto] Princes Street that you [knew you] were getting off at the next stop.'' 

This reliance on remembered turns, landmarks, gradients, and route sequences can be interpreted through spatial cognition theory. For the target group, navigation depends heavily on non-visual sensory cues and stored mental representations of familiar routes rather than continuous visual updates of the environment. This aligns with the distinction between egocentric and allocentric navigation based on broader mental representations of spatial relationships introduced by \citet{Ruggiero2021}. The findings also support travel behaviour theory, where repeated experience, perceived confidence, and familiarity shape mobility decisions. Here, memory-based navigation is a response to inaccessible and unpredictable environments, with potential benefits from real-time data accessibility aligning with \citet{ABIDI2024e31825}.

The ability to develop or adapt cognitive abilities for navigation is also in line with \citet{Passini1990}, who, in an analysis of 18 subjects per group for different visual impairment categories, provide evidence that spatio-cognitive competence is acquirable without vision and even prior seeing, with congenitally blind participants outperforming both the sighted and blindfolded control group and people with acquired total blindness, showcasing the distinction between deficiency, inefficiency, and difference \citet{Fletcher1980}.

This compatibility with the difference option corresponds to \citet{Passini1988} stating that differences in spatial competence are explainable through information access or stress, which reflects the lack of accessible data formats and cognitive overload in modern urban navigation discussed in our work. For a more in-depth overview of spatial cognition in navigation with visual impairments, see \citet{Giudice2018}.

The concentration of service activity in the central areas shows the potential of real-time support to identify correct stops and vehicles, with P3 recalling, for peak hours, ``[...] if a bus comes, you don't know what number it is. Sometimes you could be standing in a queue or a bus shelter and your one could actually come but you can't get to the front to find out the number'', which confirms the work by \citet{Cimarolli2011} discussed in Section~\ref{sec:lit-review}. Real-time data, when integrated into adaptive navigation tools, could filter relevant information, such as announcing only the approaching service of interest to reduce the burden of travel by making dynamic changes easier to manage.

Participants express both curiosity and hesitation regarding AI, agreeing with work on technology adoption by \citet{wong2025olderadults}, but contrary to our pre-analysis assumptions, participants do not express heightened paranoia, scepticism, or distrust. Our analysis indicates that familiarity with technology plays a key role in shaping attitudes towards future advancements, as participants stress the importance of assistive technology complementing, rather than overwhelm, sensory interdependence.

While these technological aids offer opportunities to enhance accessibility, our study highlights the need for tools that go beyond route guidance. P1 imagines technology that is personalisable, learning routes over time and only announcing relevant interchanges or stops via non-intrusive audio cues. Profile-based machine learning models, as described by \citet{ZHANG2024102413}, could improve reliability over time. Additionally, such systems should consider supporting haptic or selective auditory information to complement existing sensory inputs \citep{ABIDI2024e31825}. Our results also support the shift from infrastructure-based aids such as smart crossings or beacons to more context-aware, personalised, and mobile solutions \citep{MA2023101808}.

Broader limitations of this work include the sample size as well as a geographical focus on Edinburgh, which should be taken into account for generalisations and requires further research across multiple urban areas. More specifically, for accessibility, the lack of direct information for public transport vehicles and stops, previously described in detail in Section~\ref{sec:limitations}, is a frequent problem in related research. Measures are, therefore, often based on available geospatial information, for example \citet{Fransen2015} exploring ``spatial exclusion'' in the public transport system of Flanders, Belgium to address accessibility, or \citet{Dianin2024} and \citet{Lee2018} analysing ``space-time accessibility'' in the M{\"u}hlwald municipality of South Tyrol, Italy and the city of Columbus, Ohio in the United States, respectively.

\citet{Kwan2010} discuss the conventional use of spatial information, stressing the rising importance of information technology for individual accessibility. Given the previously described possibilities to integrate real-time information, we strongly recommend the collection and storage of vehicle and stop accessibility information by public transport providers to combat this pervasive challenge. The availability of, for example, assistance options for vehicles and physical stop accessibility, would provide an additional dimension for this kind of analysis.

Follow-up work in this context could benefit from larger and more diverse participant groups through extended data collection periods. Additionally, access to more granular live transport data would support a deeper analysis of the key barriers This information is already present in the internal system feeding into the live displays at select stops, as well as the corresponding Bus \& Tram App by Lothian Buses and Edinburgh Trams. Participants describe the challenge of a lack of live information in an accessible format and confirm their use of mobile apps as aids. This could be addressed with a targeted app using these existing databases, both for GPS data to identify nearby stops and vehicle data for delays and route changes, which would come at a small cost compared to physical accessible displays that necessitate retrofitting stops.

Future research should also consider older adults with multiple sensory impairments, for example vision and hearing. This is particularly relevant as the target group often relies on auditory cues and hearing-assistive technologies, such as induction hearing loops, which may be affected by changing digital infrastructure and potential electromagnetic interference from wireless devices, smartphones, electric vehicles, and other smart city urban technologies \citep{Wittich2015, ZhangYue2024, Dong2025}.


\section{Conclusion}
\label{sec:conclusion}

This study applies a mixed-methods approach to investigate how AI-enabled navigation systems and real-time data can address urban accessibility challenges and enhance independent navigation for visually impaired elderly individuals in Edinburgh. Our quantitative analysis of TfE API transport data demonstrates the structure of a highly centralised network, while the subsequent qualitative analysis of semi-structured interviews enriches this understanding through personal experiences and individual challenges. 

Our combined findings highlight the critical role of urban development and designing for accessibility. In doing so, this work contributes to a growing area of research concerning inclusive design and the development of AI solutions in urban contexts, extending existing work on accessibility and inclusivity \citep{universal-design, ASIEDUAMPEM2024101935, EISENBERG2024104837}. In the context of Edinburgh, our research offers a theoretical understanding of how service distribution patterns and infrastructure centralisation, affected by current redevelopment initiatives, contribute to social exclusion.

In terms of practical contributions, this study also offers implications for urban planners, transport agencies, and technology developers. Findings from our quantitative analysis provide insights into transport coverage and service planning of disruptions in underserved areas, while our qualitative findings emphasise the importance of personalisation and simplicity in AI tools for elderly users with vision impairments. 

Based on our findings, we recommend developing adaptive, user-centred aids that learn and respond to individuals' routes and sensory preferences, integrating multimodal interfaces such as audio and haptic feedback. Transport agencies should also prioritise accessible real-time travel information at stops and hubs, while designers and policy-makers need to involve users directly in the design of future mobility technologies.

\section*{Acknowledgements}

We would like to express our thanks to North East Sensory Services, the Royal National Institute of Blind People, Lothian Buses, and Edinburgh Trams for their willingness to assist in this research, including help in reaching participants and providing the required datasets. We are also grateful to the participants who generously gave their time and shared their experiences and challenges. This research did not receive any specific grant from funding agencies in the public, commercial, or not-for-profit sectors.

\section*{Declaration of interest}

The authors declare no conflicts of interest.

\section*{Data availability statement}

The data on Edinburgh's public transport system used in the quantitative part of this work is publicly available through the API of Transport for Edinburgh at \url{https://tfe-opendata.readme.io}. The raw qualitative dataset for interviews with participants from a group with multiple vulnerabilities is subject to participant consent as well as strict data processing and retention policies and, as such, cannot be shared.

\begin{footnotesize}

\bibliographystyle{apalike}
\bibliography{references}

\begin{thebibliography}{}

\bibitem[Aberle et~al., 2025]{ABERLE2025104140}
Aberle, C., Daubitz, S., Schwedes, O., and Gertz, C. (2025).
\newblock Measuring transport poverty with a mixed-methods approach. a comparative case study of the {German} cities {Berlin} and {Hamburg}.
\newblock {\em Journal of Transport Geography}, 125:104140.

\bibitem[Abidi et~al., 2024]{ABIDI2024e31825}
Abidi, M.~H., {Noor Siddiquee}, A., Alkhalefah, H., and Srivastava, V. (2024).
\newblock A comprehensive review of navigation systems for visually impaired individuals.
\newblock {\em Heliyon}, 10(11):e31825.

\bibitem[Ahmed, 2024]{AHMED2024100662}
Ahmed, S.~K. (2024).
\newblock How to choose a sampling technique and determine sample size for research: A simplified guide for researchers.
\newblock {\em Oral Oncology Reports}, 12:100662.

\bibitem[Akram, 2021]{Akram2021}
Akram, O. (2021).
\newblock Getting extreme poverty narrated: Methodological challenges of interviewing older persons.
\newblock {\em International Journal of Qualitative Methods}, page 16094069211016716.

\bibitem[{Al Shehri} et~al., 2022]{Shehri2022}
{Al Shehri}, W., Almalki, J., Alshahrani, S.~M., Alammari, A., Khan, F., and Alangari, S. (2022).
\newblock Assistive technology acceptance for visually impaired individuals: A case study of students in {Saudi Arabia}.
\newblock {\em PeerJ Computer Science}, 8:e886.

\bibitem[Asiedu-Ampem et~al., 2024]{ASIEDUAMPEM2024101935}
Asiedu-Ampem, G., Danso, A.~K., Ayarkwa, J., Obeng-Atuah, D., Tudzi, E.~P., and Afful, A.~E. (2024).
\newblock Barriers to accessibility of urban roads by persons with disabilities: A review of the literature.
\newblock {\em Journal of Transport and Health}, 39:101935.

\bibitem[Baddeley et~al., 2015]{Baddeley2015}
Baddeley, A., Rubak, E., and Turner, R. (2015).
\newblock {\em {Spatial point patterns: Methodology and applications with R}}.
\newblock London: Chapman \& Hall, 1 edition.

\bibitem[Barnes et~al., 2022]{Barnes2022}
Barnes, T.~L., Ahuja, M., MacLeod, S., Tkatch, R., Albright, L., Schaeffer, J.~A., and Yeh, C.~S. (2022).
\newblock Loneliness, social isolation, and all-cause mortality in a large sample of older adults.
\newblock {\em Journal of Aging and Health}, 34(6--8):883--892.

\bibitem[Batty et~al., 2012]{Batty20121009}
Batty, M., Axhausen, K.~W., Giannotti, F., Pozdnoukhov, A., Bazzani, A., Wachowicz, M., Ouzounis, G., and Portugali, Y. (2012).
\newblock Smart cities of the future.
\newblock {\em The European Physical Journal Special Topics}, 214:481--518.

\bibitem[Braun et~al., 2017]{braun-clarke2017}
Braun, V., Clarke, V., Willig, C., Rogers, W., Terry, G., and Hayfield, N. (2017).
\newblock {\em Thematic analysis}, chapter~2, pages 17--36.
\newblock London: Sage Publishing.

\bibitem[Campisi et~al., 2021]{Campisi2021}
Campisi, T., Ignaccolo, M., Inturri, G., Tesoriere, G., and Torrisi, V. (2021).
\newblock Evaluation of walkability and mobility requirements of visually impaired people in urban spaces.
\newblock {\em Research in Transportation Business and Management}, 40:100592.

\bibitem[Chemnad and Othman, 2024]{Chemnad2024}
Chemnad, K. and Othman, A. (2024).
\newblock Digital accessibility in the era of artificial intelligence -- bibliometric analysis and systematic review.
\newblock {\em Frontiers in Artificial Intelligence}, 7.

\bibitem[Chen et~al., 2024]{CHEN2024110165}
Chen, Y., Tan, P., Li, M., Yin, H., and Tang, R. (2024).
\newblock K-means clustering method based on nearest-neighbor density matrix for customer electricity behavior analysis.
\newblock {\em International Journal of Electrical Power and Energy Systems}, 161:110165.

\bibitem[Cimarolli et~al., 2012]{Cimarolli2011}
Cimarolli, V.~R., Boerner, K., Brennan-Ing, M., Reinhardt, J.~P., and Horowitz, A. (2012).
\newblock Challenges faced by older adults with vision loss: A qualitative study with implications for rehabilitation.
\newblock {\em Clinical Rehabilitation}, 26(8):748--757.

\bibitem[Clark and Evans, 1954]{Clark1954}
Clark, P.~J. and Evans, F.~C. (1954).
\newblock Distance to nearest neighbor as a measure of spatial relationships in populations.
\newblock {\em Ecology}, 35(4):445--453.

\bibitem[Cooke and Behrens, 2017]{COOKE20173003}
Cooke, S. and Behrens, R. (2017).
\newblock Correlation or cause? the limitations of population density as an indicator for public transport viability in the context of a rapidly growing developing city.
\newblock {\em Transportation Research Procedia}, 25:3003--3016.

\bibitem[Dankwa-Mullan, 2024]{Dankwa-Mullan2024}
Dankwa-Mullan, I. (2024).
\newblock Health equity and ethical considerations in using artificial intelligence in public health and medicine.
\newblock {\em Preventing Chronic Disease}, 21:240245.

\bibitem[Deverell et~al., 2022]{Deverell03042022}
Deverell, L., Bhowmik, J., Lau, B.~T., Mahmud, A.~A., Sukunesan, S., Islam, F. M.~A., McCarthy, C., and and, D.~M. (2022).
\newblock Use of technology by orientation and mobility professionals in {A}ustralia and {M}alaysia before {COVID-19}.
\newblock {\em Disability and Rehabilitation: Assistive Technology}, 17(3):260--267.

\bibitem[Dianin et~al., 2024]{Dianin2024}
Dianin, A., Gidam, M., Hauger, G., and Ravazzoli, E. (2024).
\newblock Measuring public transport accessibility to fixed activities and discretionary opportunities: A space–time approach.
\newblock {\em European Transport Research Review}, 16:9.

\bibitem[Docherty and McKiernan, 2008]{Docherty2008}
Docherty, I. and McKiernan, P. (2008).
\newblock Scenario planning for the {E}dinburgh city region.
\newblock {\em Environment and Planning C: Government and Policy}, 26(5):c0665r.

\bibitem[Dong et~al., 2025]{Dong2025}
Dong, X.-W., Qian, Y.-D., and Lu, M. (2025).
\newblock {Electromagnetic exposure levels of electric vehicle drive motors to cochlear implanted passenger}.
\newblock {\em PLoS One. 2025 May}, 20(5).

\bibitem[Dwivedi et~al., 2023]{DWIVEDI2023122579}
Dwivedi, Y.~K., Sharma, A., Rana, N.~P., Giannakis, M., Goel, P., and Dutot, V. (2023).
\newblock Evolution of artificial intelligence research in technological forecasting and social change: Research topics, trends, and future directions.
\newblock {\em Technological Forecasting and Social Change}, 192:122579.

\bibitem[{EHSCP}, 2022]{EHSCP2022}
{EHSCP} (2022).
\newblock Access and persons with disabilities in urban areas.
\newblock Technical report, {Edinburgh Health and Social Care Partnership}, Edinburgh, UK.

\bibitem[Eisenberg et~al., 2024]{EISENBERG2024104837}
Eisenberg, Y., Heider, A., Labbe, D., Gould, R., and Jones, R. (2024).
\newblock Planning accessible cities: Lessons from high quality barrier removal plans.
\newblock {\em Cities}, 148:104837.

\bibitem[Elassy et~al., 2024]{ELASSY2024100252}
Elassy, M., Al-Hattab, M., Takruri, M., and Badawi, S. (2024).
\newblock Intelligent transportation systems for sustainable smart cities.
\newblock {\em Transportation Engineering}, 16:100252.

\bibitem[Ellard-Gray et~al., 2015]{Ellard2015}
Ellard-Gray, A., Jeffrey, N.~K., Choubak, M., and Crann, S.~E. (2015).
\newblock Finding the hidden participant: Solutions for recruiting hidden, hard-to-reach, and vulnerable populations.
\newblock {\em International Journal of Qualitative Methods}, 14(5):1609406915621420.

\bibitem[Emara, 2025]{Emara2025}
Emara, I. (2025).
\newblock ``talking the same language'': The influence of sharing visual impairment identity between researchers and participants on enhancing participant recruitment and fostering rapport during interviews with blind individuals.
\newblock {\em International Journal of Qualitative Methods}, 24.

\bibitem[Ergenc and Yuksekkaya, 2024]{Ergenc15032024}
Ergenc, C. and Yuksekkaya, O. (2024).
\newblock Institutionalizing authoritarian urbanism and the centralization of urban decision-making.
\newblock {\em Territory, Politics, Governance}, 12(3):410--429.

\bibitem[Fereday and Muir-Cochrane, 2006]{Fereday2006}
Fereday, J. and Muir-Cochrane, E. (2006).
\newblock Demonstrating rigor using thematic analysis: A hybrid approach of inductive and deductive coding and theme development.
\newblock {\em International Journal of Qualitative Methods}, 5(1):80--92.

\bibitem[Fletcher, 1980]{Fletcher1980}
Fletcher, J.~F. (1980).
\newblock Spatial representation in blind children. 1: Development compared to sighted children.
\newblock {\em Journal of Visual Impairment and Blindness}, 74(10):381--385.

\bibitem[Fransen et~al., 2015]{Fransen2015}
Fransen, K., Neutens, T., Farber, S., {De Maeyer}, P., Deruyter, G., and Witlox, F. (2015).
\newblock Identifying public transport gaps using time-dependent accessibility levels.
\newblock {\em Journal of Transport Geography}, 48:176--187.

\bibitem[Fraser et~al., 2019]{Fraser2019}
Fraser, S., Beeman, I., Southall, K., and Wittich, W. (2019).
\newblock Stereotyping as a barrier to the social participation of older adults with low vision: A qualitative focus group study.
\newblock {\em BMJ Open}, 9(9):e029940.

\bibitem[Gao, 2024]{GaoEdiFestival}
Gao, X. (2024).
\newblock Research on the impact of cultural festivals on urban regeneration -- a case study of {E}dinburgh as a festival city.
\newblock {\em Highlights in Art and Design}, 6(3):1--5.

\bibitem[Giudice, 2018]{Giudice2018}
Giudice, N.~A. (2018).
\newblock Chapter 15: Navigating without vision: Principles of blind spatial cognition.
\newblock In {\em Handbook of Behavioral and Cognitive Geography}. Cheltenham: Edward Elgar Publishing.

\bibitem[Gorman et~al., 2003]{GORMAN200315}
Gorman, D., Douglas, M.~J., Conway, L., Noble, P., and Hanlon, P. (2003).
\newblock Transport policy and health inequalities: A health impact assessment of edinburgh's transport policy.
\newblock {\em Public Health}, 117(1):15--24.

\bibitem[Halden, 2002]{HALDEN2002313}
Halden, D. (2002).
\newblock Using accessibility measures to integrate land use and transport policy in {E}dinburgh and the {L}othians.
\newblock {\em Transport Policy}, 9(4):313--324.

\bibitem[Karou and Hull, 2014]{KAROU20141}
Karou, S. and Hull, A. (2014).
\newblock Accessibility modelling: Predicting the impact of planned transport infrastructure on accessibility patterns in {E}dinburgh, {UK}.
\newblock {\em Journal of Transport Geography}, 35:1--11.

\bibitem[Kirkpatrick, 2015]{IETneatebox}
Kirkpatrick, K. (2015).
\newblock News: Using technology to help people.
\newblock {\em Communications of the ACM}, Volume 58(2):21--23.

\bibitem[Kolotouchkina et~al., 2022]{KOLOTOUCHKINA2022103613}
Kolotouchkina, O., Barroso, C.~L., and Sánchez, J. L.~M. (2022).
\newblock Smart cities, the digital divide, and people with disabilities.
\newblock {\em Cities}, 123:103613.

\bibitem[K\"{o}me\c{c}li, 2024]{su16062485}
K\"{o}me\c{c}li, P. (2024).
\newblock Accessibility of urban tourism in historical areas: Analysis of {UNESCO} {W}orld {H}eritage sites in {S}afranbolu.
\newblock {\em Sustainability}, 16(6):2485.

\bibitem[Kulkarni et~al., 2023]{kulkarniBusStopCVRealtimeAI2023}
Kulkarni, M., Li, C., Ahn, J.~J., Ma, K. O.~Y., Zhang, Z., Saugstad, M., Wu, K., Eisenberg, Y., et~al. (2023).
\newblock Busstopcv: A real-time ai assistant for labeling bus stop accessibility features in streetscape imagery.
\newblock In {\em Proceedings of the 25th International ACM SIGACCESS Conference on Computers and Accessibility}, pages 1--6. New York: Association for Computing Machinery.

\bibitem[Kwan and Weber, 2010]{Kwan2010}
Kwan, M.-P. and Weber, J. (2010).
\newblock Individual accessibility revisited: Implications for geographical analysis in the twenty-first century.
\newblock {\em Geographical Analysis}, 35(4):341--353.

\bibitem[Lee and Miller, 2018]{Lee2018}
Lee, J. and Miller, H.~J. (2018).
\newblock Measuring the impacts of new public transit services on space-time accessibility: An analysis of transit system redesign and new bus rapid transit in {C}olumbus, {Ohio}, {USA}.
\newblock {\em Applied Geography}, 93:47--63.

\bibitem[Lloyd, 1982]{Lloyd1982}
Lloyd, S. (1982).
\newblock Least squares quantization in pcm.
\newblock {\em IEEE Transactions on Information Theory}, 28(2):129--137.

\bibitem[Luo et~al., 2019]{LUO2019102505}
Luo, D., Cats, O., {van Lint}, H., and Currie, G. (2019).
\newblock Integrating network science and public transport accessibility analysis for comparative assessment.
\newblock {\em Journal of Transport Geography}, 80:102505.

\bibitem[Ma et~al., 2023]{MA2023101808}
Ma, B., Yang, J., Wong, F. K.~Y., Wong, A. K.~C., Ma, T., Meng, J., Zhao, Y., Wang, Y., et~al. (2023).
\newblock Artificial intelligence in elderly healthcare: A scoping review.
\newblock {\em Ageing Research Reviews}, 83:101808.

\bibitem[MacQueen, 1967]{MacQueen1967}
MacQueen, J.~B. (1967).
\newblock Some methods for classification and analysis of multivariate observations.
\newblock In {\em Proceedings of the 5th Berkeley Symposium on Mathematical Statistics and Probability}, volume~1, pages 281--297.

\bibitem[McMullin, 2023]{McMullin2023}
McMullin, C. (2023).
\newblock Transcription and qualtitative methods: Implications for third sector research.
\newblock {\em VOLUNTAS: International Journal of Voluntary and Nonprofit Organizations}, 34:140--153.

\bibitem[Meneghetti et~al., 2021]{Meneghetti2021}
Meneghetti, C., Miola, L., Toffalini, E., Pastore, M., and Pazzaglia, F. (2021).
\newblock Learning from navigation, and tasks assessing its accuracy: The role of visuospatial abilities and wayfinding inclinations.
\newblock {\em Journal of Environmental Psychology}, 75:101614.

\bibitem[Miranda et~al., 2020]{Miranda2020}
Miranda, E., {Batista e Silva}, J., and {da Costa}, A.~R. (2020).
\newblock Emergence and structure of urban ventralities in a medium-sized historic city.
\newblock {\em SAGE Open}, 10(3):2158244020930002.

\bibitem[Montarzino et~al., 2007]{Montarzino01012007}
Montarzino, A., Robertson, B., Aspinall, P., Ambrecht, A., Findlay, C., Hine, J., and Dhillon, B. (2007).
\newblock The impact of mobility and public transport on the independence of visually impaired people.
\newblock {\em Visual Impairment Research}, 9(2-3):67--82.

\bibitem[Muhsin et~al., 2024]{Mushin2024}
Muhsin, Z.~J., Qahwaji, R., Ghanchi, F., and Al-Taee, M. (2024).
\newblock Review of substitutive assistive tools and technologies for people with visual impairments: recent advancements and prospects.
\newblock {\em Journal on Multimodal User Interfaces}, 18:135--156.

\bibitem[Nykiforuk et~al., 2021]{Nykiforuk2021}
Nykiforuk, C.~I., Glenn, N.~M., Hosler, I., Craig, H., Reynard, D., Molner, B., Candlish, J., and Lowe, S. (2021).
\newblock Understanding urban accessibility: A community-engaged pilot study of entrance features.
\newblock {\em Social Science and Medicine}, 273:113775.

\bibitem[Padeiro et~al., 2022]{Padeiro2021}
Padeiro, M., de~S{\~a}o~Jos{\'e'}, J., Amado, C., Sousa, L., Oliveira, C.~R., Esteves, A., and McGarrigle, J. (2022).
\newblock Neighborhood attributes and well-being among older adults in urban areas: A mixed-methods systematic review.
\newblock {\em Research on Ageing}, 44(5-6):351--368.

\bibitem[Parady et~al., 2023]{PARADY2023344}
Parady, G., Suzuki, K., Oyama, Y., and Chikaraishi, M. (2023).
\newblock Activity detection with google maps location history data: Factors affecting joint activity detection probability and its potential application on real social networks.
\newblock {\em Travel Behaviour and Society}, 30:344--357.

\bibitem[Parzen, 1962]{Parzen1962}
Parzen, E. (1962).
\newblock On estimation of a probability density function and mode.
\newblock {\em The Annals of Mathematical Statistics}, 33(3):1065--1076.

\bibitem[Passini and Proulx, 1988]{Passini1988}
Passini, R. and Proulx, G. (1988).
\newblock Wayfinding without vision: An experiment with congenitally totally blind people.
\newblock {\em Environment and Behavior}, 20(2):227--252.

\bibitem[Passini et~al., 1990]{Passini1990}
Passini, R., Proulx, G., and Rainville, C. (1990).
\newblock The spatio-cognitive abilities of the visually impaired population.
\newblock {\em Environment and Behavior}, 22(1):91--118.

\bibitem[Pineda, 2024]{Pineda2024}
Pineda, V.~S. (2024).
\newblock {\em How cities shape our experience}, pages 61--84.
\newblock Singapore: Springer Nature.

\bibitem[Prandi et~al., 2023]{prandiAccessibleWayfindingNavigation2023}
Prandi, C., Barricelli, B.~R., Mirri, S., and Fogli, D. (2023).
\newblock Accessible wayfinding and navigation: A systematic mapping study.
\newblock {\em Universal Access in the Information Society}, 22(1):185--212.

\bibitem[Proudfoot, 2023]{Proudfoot2023}
Proudfoot, K. (2023).
\newblock Inductive/deductive hybrid thematic analysis in mixed methods research.
\newblock {\em Journal of Mixed Methods Research}, 17(3):308--326.

\bibitem[Rasouli~Kahaki et~al., 2023]{rasoulikahakiCane2023}
Rasouli~Kahaki, Z., Karimi, M., Taherian, M., and Simi, R. (2023).
\newblock Development and validation of a white cane use perceived advantages and disadvantages {(WCPAD)} questionnaire.
\newblock {\em BMC Psychology}, 11(1):253.

\bibitem[Remillard et~al., 2023]{Remillard2023}
Remillard, E.~T., Koon, L.~M., Mitzner, T.~L., and Rogers, W.~A. (2023).
\newblock Everyday challenges for individuals aging with vision impairment: Technology implications.
\newblock {\em The Gerontologist}, 64(6):gnad169.
\newblock [Accessed: 29 November 2024].

\bibitem[Ripley, 1976]{Ripley1976}
Ripley, B.~D. (1976).
\newblock {The second-order analysis of stationary point processes}.
\newblock {\em Journal of Applied Probability}, 13(2):255--266.

\bibitem[Ripley, 1977]{Ripley1977}
Ripley, B.~D. (1977).
\newblock {Modelling spatial patterns}.
\newblock {\em Journal of the Royal Statistical Society: Series B}, 39(2):172--192.

\bibitem[Rosenblatt, 1956]{Rosenblatt1956}
Rosenblatt, M. (1956).
\newblock Remarks on some nonparametric estimates of a density function.
\newblock {\em The Annals of Mathematical Statistics}, 27(3):832--837.

\bibitem[Ruggiero et~al., 2022]{Ruggiero2021}
Ruggiero, G., Ruotolo, F., and Iachini, T. (2022).
\newblock How ageing and blindness affect egocentric and allocentric spatial memory.
\newblock {\em Quarterly Journal of Experimental Psychology}, 75(9):1628--1642.

\bibitem[Schinazi et~al., 2015]{Schinazi2015}
Schinazi, V.~R., Thrash, T., and Chebat, D.-R. (2015).
\newblock Spatial navigation by congenitally blind individuals.
\newblock {\em WIREs Cognitive Science}, 7(1):37--58.

\bibitem[Schulz et~al., 2015]{Schulz2015}
Schulz, R., Wahl, H.-W., Matthews, J.~T., Dabbs, A. D.~V., Beach, S.~R., and Czaja, S.~J. (2015).
\newblock Advancing the aging and technology agenda in gerontology.
\newblock {\em The Gerontologist}, 55(5):724--734.

\bibitem[Silverman, 1986]{Silverman1986}
Silverman, B.~W. (1986).
\newblock {\em Density estimation for statistics and data analysis}.
\newblock London: Chapman \& Hall.

\bibitem[Singh et~al., 2024]{SmartStickResearch}
Singh, A.~K., Rajarajeshwari, P., Rathika, R., Shobiya, B., Keerthi, P., and Marieswaran, M. (2024).
\newblock Common smart stick for blind and elderly people to detect environmental factors and free navigation.
\newblock In Singh, B.~K., Sinha, G., and Pandey, R., editors, {\em Biomedical Engineering Science and Technology}, pages 388--400. Cham: Springer Nature Switzerland.

\bibitem[Sol{\'a} et~al., 2018]{GILSOLA20181}
Sol{\'a}, A.~G., Vilhelmson, B., and Larsson, A. (2018).
\newblock Understanding sustainable accessibility in urban planning: Themes of consensus, themes of tension.
\newblock {\em Journal of Transport Geography}, 70:1--10.

\bibitem[Steinfeld and Maisel, 2012]{universal-design}
Steinfeld, E. and Maisel, J. (2012).
\newblock {\em Universal design : Creating inclusive environments}, chapter 2: Defining universal design.
\newblock Hoboken, NJ, USA: Wiley \& Sons.

\bibitem[Stratton, 2024]{Stratton2024}
Stratton, S.~J. (2024).
\newblock urposeful sampling: Advantages and pitfalls.
\newblock {\em Prehosp Disaster Med}, 39:21--122.

\bibitem[Sun et~al., 2020]{Sun20200921}
Sun, L., Chen, J., Li, Q., and Huang, D. (2020).
\newblock Dramatic uneven urbanization of large cities throughout the world in recent decades.
\newblock {\em Nature Communications}, 11(5366).

\bibitem[Thomas and Harden, 2008]{Thomas2008}
Thomas, J. and Harden, A. (2008).
\newblock Methods for the thematic synthesis of qualitative research in systematic reviews.
\newblock {\em BMC Medical Research Methodology}, 8:45.

\bibitem[{United Nations}, 2016]{UN-SUD}
{United Nations} (2016).
\newblock Good practices of accessible urban development: making urban environments inclusive and fully accessible to all.
\newblock Technical Report ST/ESA/364, {Department of Economic and Social Affairs of the United Nations (DESA)}, New York City, NY, USA.

\bibitem[{United Nations}, 2023]{UNWSR}
{United Nations} (2023).
\newblock World social report 2023: Leaving no one behind in an ageing world.
\newblock Technical Report ST/ESA/379, Department of Economic and Social Affairs of the United Nations (DESA), New York City, NY, USA.

\bibitem[{Van Acker} et~al., 2010]{VanAcker01032010}
{Van Acker}, V., Wee, B.~V., and Witlox, F. (2010).
\newblock When transport geography meets social psychology: Toward a conceptual model of travel behaviour.
\newblock {\em Transport Reviews}, 30(2):219--240.

\bibitem[Vincent and Hartt, 2024]{Vincent2024}
Vincent, E. and Hartt, M. (2024).
\newblock Ageing and the sensory city.
\newblock {\em Town Planning Review}, 95(2):175--195.

\bibitem[Wahl, 2013]{WernerWahl2013}
Wahl, H.-W. (2013).
\newblock The psychological challenge of late-life vision impairment: Concepts, findings, and practical implications.
\newblock {\em Journal of Ophthalmology}, page 278135.

\bibitem[Wang and Loo, 2024]{WANG2024103983}
Wang, H. and Loo, B. P.~Y. (2024).
\newblock The public transport disadvantaged in a highly transit-oriented city: An analytical framework, key challenges and opportunities.
\newblock {\em Journal of Transport Geography}, 120:103983.

\bibitem[Wang et~al., 2022]{WANG2022103611}
Wang, S., Yung, E. H.~K., and Sun, Y. (2022).
\newblock Effects of open space accessibility and quality on older adults' visit: Planning towards equal right to the city.
\newblock {\em Cities}, 125:103611.

\bibitem[Whitmarsh, 2005]{Whitmarsh01012005}
Whitmarsh, L. (2005).
\newblock The benefits of guide dog ownership.
\newblock {\em Visual Impairment Research}, 7(1):27--42.

\bibitem[Wittich et~al., 2015]{Wittich2015}
Wittich, W., Southall, K., and Johnson, A. (2015).
\newblock {Usability of assistive listening devices by older adults with low vision}.
\newblock {\em Disability and Rehabilitation: Assistive Technology}, 11:564–--571.

\bibitem[Wong et~al., 2025]{wong2025olderadults}
Wong, A. K.~C., Lee, J. H.~T., Zhao, Y., Lu, Q., Yang, S., and Hui, V. C.~C. (2025).
\newblock Exploring older adults' perspectives and acceptance of ai-driven health technologies: Qualitative study.
\newblock {\em JMIR Ageing}, 8(e66778).

\bibitem[Wong, 2018]{WONG2018300}
Wong, S. (2018).
\newblock The limitations of using activity space measurements for representing the mobilities of individuals with visual impairment: A mixed methods case study in the {San Francisco Bay Area}.
\newblock {\em Journal of Transport Geography}, 66:300--308.

\bibitem[Wood et~al., 2022]{Wood2022}
Wood, G. E.~R., Pykett, J., Daw, P., Agyapong-Badu, S., Banchoff, A., King, A.~C., and Stathi, A. (2022).
\newblock The role of urban evironments in promoting active and healthy aging: A systematic scoping review of citizen science approaches.
\newblock {\em Journal of Urban Health}, 99(3):427--456.

\bibitem[{World Health Organization}, 1998]{WHOQoL}
{World Health Organization} (1998).
\newblock Programme on mental health : {WHOQOL} user manual.
\newblock Technical Report WHO/HIS/HSI Rev.2012.03, Division of Mental Heath and Prevention of Substance Abuse of the World Health Organization.

\bibitem[Yu et~al., 2014]{YU201480}
Yu, H., Liu, P., Chen, J., and Wang, H. (2014).
\newblock Comparative analysis of the spatial analysis methods for hotspot identification.
\newblock {\em Accident Analysis and Prevention}, 66:80--88.

\bibitem[Zhang and Yue, 2024]{ZhangYue2024}
Zhang, M. and Yue, X. (2024).
\newblock A study on audio-frequency near-field electromagnetic interference system in wearable audio devices.
\newblock In {\em 2024 IEEE International Symposium on Electromagnetic Compatibility, Signal and Power Integrity (EMC+SIPI)}, pages 624--629.

\bibitem[Zhang et~al., 2024]{ZHANG2024102413}
Zhang, S., Luo, Z., Yang, L., Teng, F., and Li, T. (2024).
\newblock A survey of route recommendations: Methods, applications, and opportunities.
\newblock {\em Information Fusion}, 108:102413.

\end{thebibliography}

\end{footnotesize}

\end{document}